\documentclass[aps,prb,preprint,superscriptaddress,final,preprintnumbers,showpacs,showkeys]{revtex4-1}
\usepackage{graphicx}
\usepackage{color}
\usepackage{rotating}
\usepackage{amssymb}
\usepackage[colorlinks=true,citecolor=blue]{hyperref}
\usepackage{amsmath}
\usepackage{amsfonts}
\usepackage{gensymb}
\usepackage{natbib}

\begin{document}

\preprint{}

\title{Electronic and vibrational properties of V$_2$C-based MXenes: from experiments to first-principles modeling}

\author{Aur\'elie Champagne}
\affiliation{Institute of Condensed Matter and Nanoscience (IMCN), Universit\'e catholique de Louvain, B-1348 Louvain-la-Neuve, Belgium}
\author{Lu Shi}
\affiliation{Institute of Condensed Matter and Nanoscience (IMCN), Universit\'e catholique de Louvain, B-1348 Louvain-la-Neuve, Belgium}
\affiliation{Laboratoire des Mat\'eriaux et du G\'enie Physique (LMGP), Universit\'e Grenoble Alpes, 38000 Grenoble, France}
\author{Thierry Ouisse}
\affiliation{Laboratoire des Mat\'eriaux et du G\'enie Physique (LMGP), Universit\'e Grenoble Alpes, 38000 Grenoble, France}
\author{Beno\^it Hackens}
\affiliation{Institute of Condensed Matter and Nanoscience (IMCN), Universit\'e catholique de Louvain, B-1348 Louvain-la-Neuve, Belgium}
\author{Jean-Christophe Charlier}
\affiliation{Institute of Condensed Matter and Nanoscience (IMCN), Universit\'e catholique de Louvain, B-1348 Louvain-la-Neuve, Belgium}

\date{\today}

\begin{abstract}
In the present work, the electronic and vibrational properties of both pristine V$_2$C and fully-terminated V$_2$CT$_2$ (where T = F, O, OH) 2D monolayers are investigated using density functional theory. Firstly, the atomic structures of V$_2$C-based MXene phases are optimized and their respective dynamical stabilities are discussed. Secondly, electronic band structures are computed indicating that V$_2$C is metallic as well as all the corresponding functionalized systems. Thirdly, the vibrational properties (phonon frequencies and spectra) of V$_2$C-based MXenes are computed thanks to density functional perturbation theory and reported for the first time. Both Raman (E$_g$, A$_{1g}$) and infrared active (E$_u$, A$_{2u}$) vibrational modes are predicted \textit{ab initio} with the aim to correlate the experimental Raman peaks with the calculated vibrational modes and to assign them with specific atomic motions. The effect of the terminal groups on the vibrational properties is emphasized as well as on the presence and position of the corresponding Raman peaks. Our results provide new insights for the identification and characterization of V$_2$C-based samples using Raman spectroscopy.
\end{abstract}


\maketitle

\section{\label{sec:level0}Introduction}
Since the discovery of graphene and its outstanding properties,~\cite{cite1,FNRS4bis} two-dimensional (2D) materials have attracted a major interest in the field of material science and device processing. Purely 2D crystals are a subclass of nano-materials that exhibit interesting physical characteristics due to the quantum confinement of their electrons. At present, more than dozen different 2D crystals have been reported,~\cite{FNRS6} including hexagonal boron-nitride (\textit{h}-BN),~\cite{cite2} transition metal dichalcogenides (MoS$_2$, MoSe$_2$, WS$_2$, NbSe$_2$,...),~\cite{cite4} thin oxide layers (TiO$_2$, MoO$_3$, WO$_3$,...), silicene,~\cite{ad1} phosphorene,~\cite{cite3} germanene,~\cite{ad2,ad3} stanene, borophene, etc. These novel 2D systems also exhibit exotic properties suggesting new possible applications. However, the lack of flexibility in their chemical composition and the weak interlayer (van der Waals) interactions limit their use. Recently, the family of 2D materials has been significantly expanded by introducing 2D layers of transition metal carbides, nitrides and carbo-nitrides, known as MXenes.~\cite{cite5,cite6,cite18,ad4} MXenes have been shown to be very promising building blocks of an impressive number of potential applications including energy storage devices,~\cite{FNRS16} such as hydrogen storage,~\cite{FNRS17} Li and multivalent ion batteries,~\cite{FNRS18,cite11,cite12,cite13,FNRS20bis} and electrochemical capacitors,~\cite{cite15,FNRS22,cite14,FNRS23,FNRS23bis,ad5} thermoelectric materials,~\cite{FNRS24} sensors,~\cite{cite7} actuators,~\cite{FNRS26,FNRS26bis}... Another exciting physical property is their extreme volumetric capacitance, exceeding 900\,F~cm$^{-3}$, while the best reported one for carbon-based electrode is around 300\,F~cm$^{-3}$.~\cite{cite13, FNRS20bis} Moreover, they exhibit good electromagnetic interference shielding abilities.~\cite{FNRS27} First-principles calculations indicate good mechanical properties, including high elastic constant,~\cite{cite8} and flexibility.~\cite{cite9} Finally, they can potentially pave the way for future spintronics devices, since their electronic and magnetic properties can be tuned via surface functionalization.~\cite{cite16,cite17}\\
MXene structures are generally produced by selectively etching layers of \textit{sp}-elements from their corresponding 3D MAX phase. The MAX phases are layered 3D solids composed of 2D sheets of MX separated by A layers,~\cite{FNRS5} thus exhibiting a general formula M$_{n+1}$AX$_n$, where M represents an early transition metal (Sc, Ti, V, Cr, Zr, Nb, Mo, Hf, Ta), A represents an element from groups 13 to 16 (Al, Si, P, Ga, Ge, As, In, Sn), X represents either a carbon (C) or a nitrogen (N) atom, and where $n$ varies from 1 to 3. In contrast with graphite-like materials with strong intralayer covalent bonds and weak van der Waals interlayer interactions, the MAX phases are mostly composed of covalent, ionic and metallic strong bonds. The chemical bonds between A and M$_{n+1}$X$_n$ are weaker than those between M-X, allowing for the extraction of A layers from the 3D crystals.~\cite{cite5} Although the removal of A layers from the MAX phases cannot be achieved with the usual mechanical exfoliation method,~\cite{cite5} M. Barsoum and coworkers demonstrated the removal of Al layers from the Ti$_3$AlC$_2$ MAX phase by hydrofluoric acid treatment and sonication,~\cite{cite5} resulting in a 3D to 2D transformation. Today, more than 20 separate MXenes have been successfully synthesized using a similar procedure, and dozens more predicted.~\cite{cite6} However, it is still presently highly challenging to exfoliate pure MXene mono-sheets from the 3D MAX phase. In addition, due to the use of etching agents, MXenes are always terminated with functional surface groups, such as $-$F, $=$O and $-$OH that are randomly distributed.~\cite{cite18} This random distribution of the terminal groups was confirmed by electron energy-loss spectroscopy in transmission electron microscopy (STEM-EELS),~\cite{ad6} nuclear magnetic resonance (NMR) spectroscopy,~\cite{FNRS22,ad8} and X-ray photo electron spectroscopy~\cite{ad8bis} for various transition metal carbides. The general formula of these chemically terminated MXene crystals is therefore M$_{n+1}$X$_n$T$_{n+1}$ where T represents the terminal groups.~\cite{cite18}\\
In the present work, the structural and electronic properties of V$_2$C-based MXenes (pristine and V$_2$CT$_2$ systems terminated with T = F, O and OH) are investigated using first-principles DFT techniques.~\cite{FNRS1,FNRS1bis} When considering a non-zero magnetic moment for the V atoms, J. Hu \textit{et al.}~\cite{FNRS20bis} found that the antiferromagnetic configuration is the most stable phase for V$_2$C, with a very small magnetic moment around 0.14\,$\mu_B$.~\cite{Add1,cite16} Given that there is no experimental evidence of magnetic properties for the V$_2$C system as confirmed by the present \textit{ab initio} simulations, only the non-magnetic phases of V$_2$C systems will be considered here. The dynamical stability is studied for different functionalization configurations, based on density functional perturbation theory (DFPT) approach.~\cite{FNRS2,FNRS2bis} Atomic structures exhibiting imaginary vibrational frequencies are noted as dynamically unstable. Both Raman (E$_{g}$ and A$_{1g}$) and infrared active (E$_u$ and A$_{2u}$) vibrational modes, which are presently missing in the literature, are predicted \textit{ab initio}. Electronic properties including electronic band structures and densities of states (DOS), as well as vibrational properties such as Raman spectra are reported and directly compared to experimental measurements. A similar investigation was recently done for Ti$_3$C$_2$-based systems,~\cite{ad9} while a stability study based on phonon frequencies investigation was done for various single layer MXene structures.~\cite{FNRS16}

\section{\label{Comput}Computational details and characterization techniques}

The structural, electronic and vibrational properties are calculated with the \textsc{Abinit} package,~\cite{FNRS3,FNRS3bis} which is based on plane-wave basis sets to represent the electronic wavefunctions and charge density. \textit{Ab initio} calculations are performed using the generalized gradient approximation (GGA) for the exchange-correlation functional, as proposed by Perdew, Burke and Ernzerhof (PBE).~\cite{66Mem} As the partially filled $d$ bands and thus localized electrons in V atoms are usually not well represented by the GGA functional, the GGA plus Hubbard U (GGA+U)~\cite{newB,newC} functional is used to predict the electronic properties with a Hubbard correction of $\sim$4\,eV for the V atoms, as also considered in previous works.~\cite{FNRS20bis,newA} Optimized norm-conserving Vanderbilt pseudopotentials (ONCVPSP-PBE)~\cite{pseudo} are used to describe the core-valence interaction. Configurations of H-$1s^1$, C-$2s^22p^2$, O-$2s^22p^4$, F-$2s^22p^5$ and V-$3s^23p^63d^34s^2$ are treated as valence electrons. A plane-wave basis set with a converged energy cutoff of 80\,Ry is used to represent the wavefunctions. The first Brillouin zone is sampled with a $18 \times 18 \times 1$ Monkhorst-Pack $k$-point grid. The atomic positions are optimized until the largest force is smaller than 2.5~10$^{-4}$\,eV$/$\AA. To prevent undesired interaction between isolated monolayers, a vacuum spacing of at least 10\,\AA$ $ is introduced. Spin-orbit coupling (SOC) effects have shown negligible influence on our computed structural and electronic properties of V$_2$C-based MXene. Hence, the following results are presented without taking SOC effects into account.\\
In order to validate the calculations, Raman measurements of exfoliated flakes are also carried out. We start sample preparation from V$_2$AlC single crystals prepared using a method reported previously.~\cite{LuShi} The flakes are synthesized by etching Al layers from V$_2$AlC by immersing the 3D crystal in diluted (40\,wt.\%) hydrofluoric acid (HF) for 72h at room temperature. Sonication is then performed in the 40wt.\% HF solution at 80$\degree$C for 8h, with the aim to separate the layered solid into flakes. The exfoliated flakes are then washed with deionized water and dried. The presence of terminal groups is evidenced by Energy Dispersive Spectroscopy (EDS) measurements, reporting an atomic ratio of 28:35:27:10 for V:C:O:F, respectively,~\cite{luunp} which is purely qualitative since quantitative measurements are not easily performed on light elements. For micro-Raman measurements, the V$_2$C-based MXene flakes are simply deposited on a glass slide and not transfered on a substrate. Raman spectroscopy is performed at room temperature with a LabRam Horiba spectrometer with a laser wavelength of 514\,nm. The laser beam is focused on the sample with a $100\cdot$ objective (NA=0.95) and the incident power is kept below 1\,mW. A grating of 2400 lines per mm is used in order to achieve a high spectral resolution.
 
\section{\label{sec:level2}Results and discussion}

\subsection{Structural properties}

\begin{figure*}
\centering
\begin{tabular}{ccccc}
    (a) & (b) & (c) & (d) & (e) \\
    \includegraphics[width=.16\textwidth]{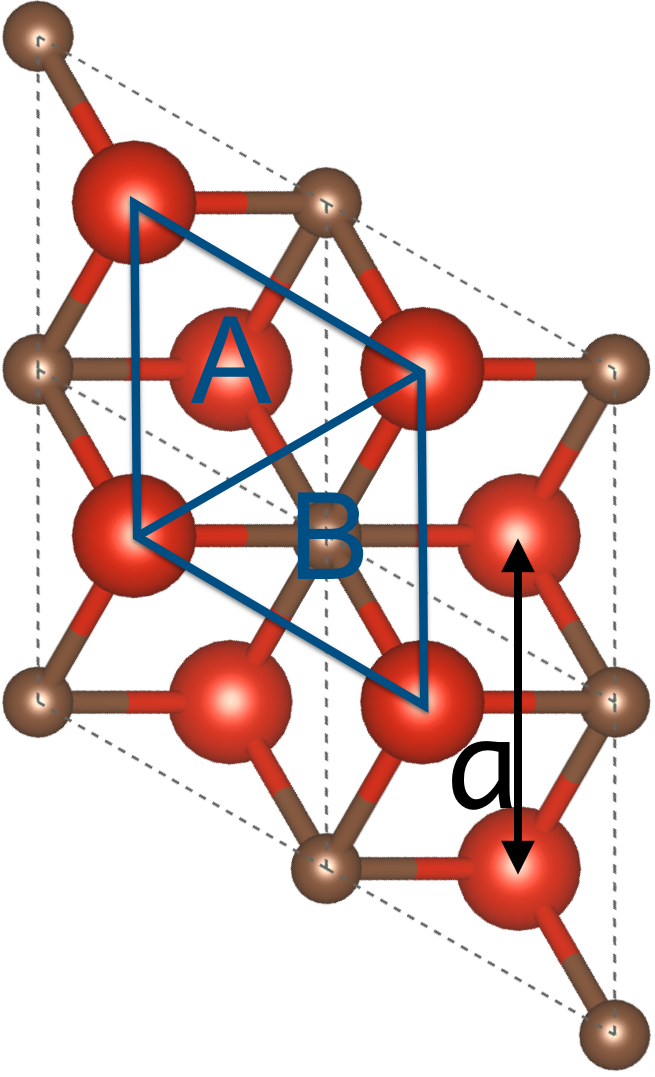} &
    \includegraphics[width=.16\textwidth]{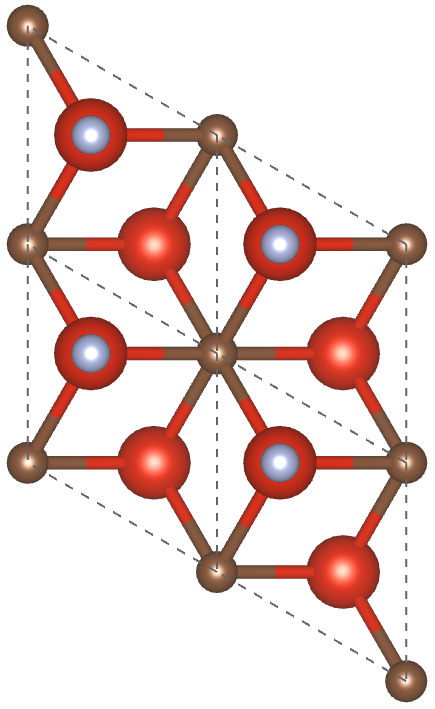} &
    \includegraphics[width=.20\textwidth]{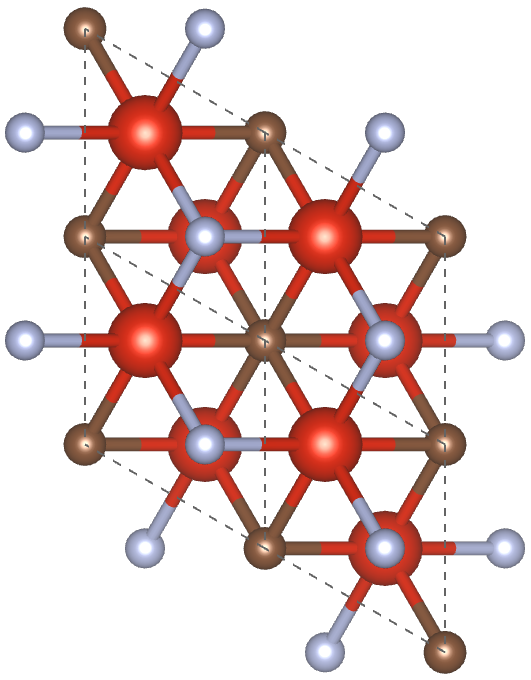} &
    \includegraphics[width=.18\textwidth]{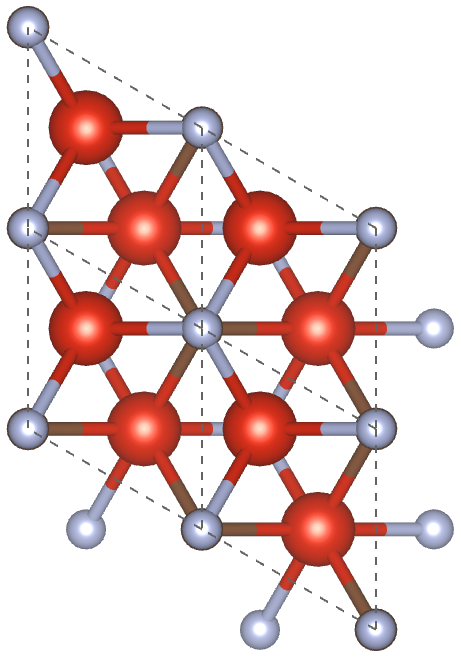} &
    \includegraphics[width=.16\textwidth]{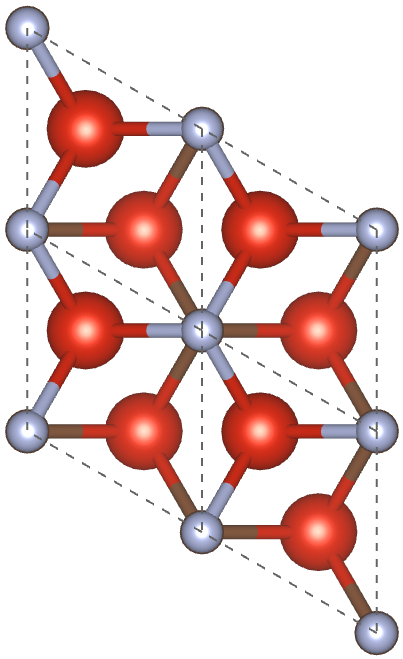} \\
    \includegraphics[width=.16\textwidth]{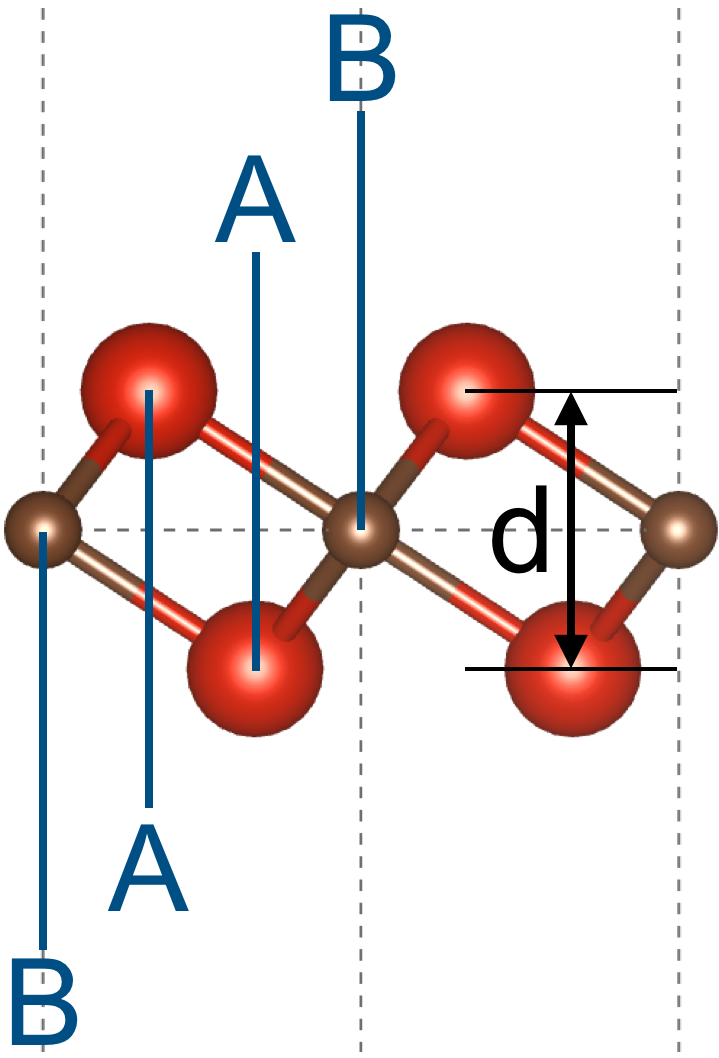} &
    \includegraphics[width=.16\textwidth]{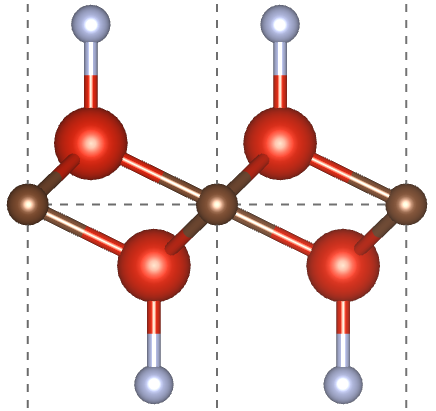} &
    \includegraphics[width=.20\textwidth]{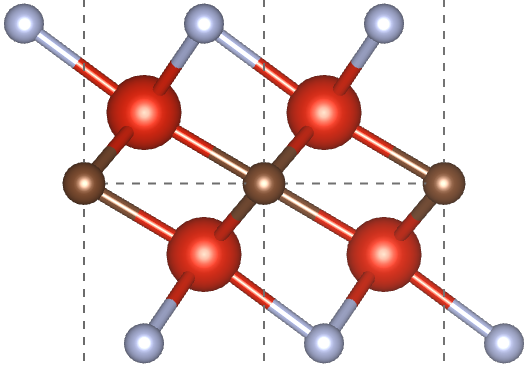} &
    \includegraphics[width=.18\textwidth]{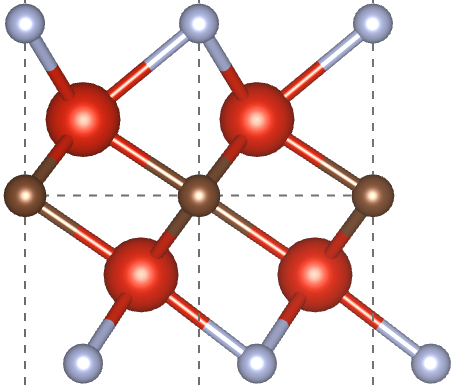} &
    \includegraphics[width=.16\textwidth]{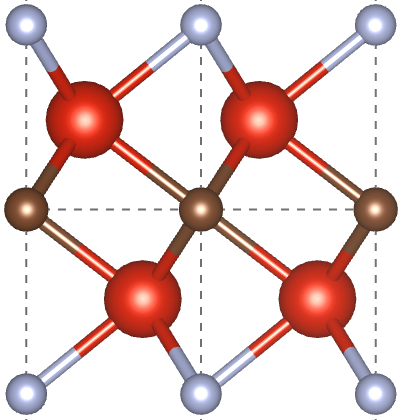} \\
\end{tabular}
\caption{(Color online) Atomic structure of (a) pristine V$_2$C and four models for terminated V$_2$CT$_2$: (b) MD1, (c) MD2, (d) MD3, and (e) MD4 - top and side views. Vanadium (V) and carbon (C) atoms are respectively in red (gray) and brown (dark gray), while the functional group (T = F, O, OH) is in blue (light gray). In (a), A and B indicate the two types of hollow sites, $a$ and $d$ are the lattice constant and the layer thickness, respectively. \label{fig:models}}
\end{figure*}

The ground state structure of the pristine V$_2$C with fully relaxed geometry is found to be hexagonal. The unit cell includes three atoms, two vanadium (V) and one carbon (C), which resides in the $P\overline{3}m1$ space group. The V atoms are located at $\left( \frac{1}{3},\frac{2}{3},z\right)$ and $\left(\frac{2}{3},\frac{1}{3},-z\right)$ on the $2d$ Wyckoff sites, and the C atom is located at $\left(0,0,0\right)$ on the $1a$ Wyckoff site. These atoms are therefore arranged in a three atomic layered structure where C layer is sandwiched between the two V layers as illustrated in Fig.~\ref{fig:models}(a). The lattice constant $a$ is equal to 2.89\,\AA$ $ and the layer thickness $d$, defined as the V-V distance, is equal to 2.18\,\AA [Fig.~\ref{fig:models}(a)].\\

Based on the unit cell of pristine V$_2$C, functionalized V$_2$CT$_2$ MXene structures with T = F, O and OH terminations are constructed. Two types of hollow sites on the surface of V$_2$C can be distinguished: A sites correspond to FCC sites for which no C atom is present under the V atoms, while B sites correspond to HCP sites located on the top of a C atom [Fig.~\ref{fig:models}(a)]. As previously proposed by Khazaei \textit{et al}.,~\cite{cite7} different functionalization models can be built according to the lattice positions of the termination groups. Four models are proposed here [Fig.~\ref{fig:models}(b-e)]. In model 1 (MD1), two functional groups of the same type (F, O or OH) are positioned on the top of the two transition metal atoms V [Fig.~\ref{fig:models}(b)]. In the case of model 2 (MD2), the two functional groups are located on the top of the hollow sites A [Fig.~\ref{fig:models}(c)]. Model 3 (MD3) considers one functional group on the top of the hollow site A and a second functional group of the same type on the top of the hollow site B [Fig.~\ref{fig:models}(d)]. At last, model 4 (MD4) presents two functional groups located on the top of the hollow sites B [Fig.~\ref{fig:models}(e)]. For each type of functionalization, the four configurations are systematically investigated and their respective stability are compared. The relative structural stabilities of the four models usually depend on the possibility for the transition metal to provide sufficient electrons to both the carbon atoms and the functional groups.~\cite{cite7} Table~\ref{tab:functionV2C} summarizes the structural parameters computed \textit{ab initio} for the four models. To ascertain the stability of the functionalized V$_2$CT$_2$ MXene mono-sheets and in order to determine the most stable configuration for each system, the formation energies are computed, according to the following formula:~\cite{ad11} \begin{equation}
\Delta \text{H}_{\text{f}} = \text{E}_{\text{tot}} \left( \text{V}_2\text{CT}_2 \right) - \text{E}_{\text{tot}} \left( \text{V}_2 \text{C} \right) - \text{E}_{\text{tot}} \left( \text{T} _2 \right)-2\mu_{\text{T}}
\end{equation}
where E$_{\text{tot}}$(V$_2$C) and E$_{\text{tot}}$(V$_2$CT$_2$) stand for the total energy of the pristine and terminated MXene, respectively. Depending on the terminal groups, E$_{\text{tot}}$(T$_2$) corresponds to the total energy of F$_2$, O$_2$ or O$_2$+H$_2$ and $\mu_{\text{T}}$ is the chemical potential of the chemisorbed groups. For all the terminated structures, the MXene surfaces are assumed to be fully-terminated with no pending bond remaining, as suggested by Khazaei \textit{et al}.,~\cite{cite7} and $\mu_{\text{T}}$ is fixed at 0.0\,eV. The large negative values reported in Table~\ref{tab:functionV2C}, corresponding to the formation energies of all V$_2$CT$_2$ systems, indicate the spontaneous and exothermic formation of strong ionic bonds between vanadium atoms and terminal groups.~\cite{cite7}.
\begin{table*}
\centering
\caption{Structural parameters and formation energies for pristine V$_2$C and MD1-MD4 functionalized systems. The ground state (GS) configuration of V$_2$CO$_2$ is also included.\label{tab:functionV2C}}
\begin{tabular}{l@{\hskip 0.5cm} l@{\hskip 0.5cm} c@{\hskip 0.3cm} c@{\hskip 0.3cm} c@{\hskip 0.3cm} c}
\hline
\hline
 & Model & Lattice parameter &  V-V distance & Thickness & Formation energy \\
 & & a (\AA) & (\AA) & d (\AA) & (eV) \\
\hline
V$_2$C & & 2.89 & 2.18 & 2.18 & 0 \\
\hline
V$_2$CF$_2$ & MD1 & 3.20 & 1.79 & 5.28 & $-$8.65 \\
 & MD2 & 2.99 & 2.04 & 4.59 & $-$9.45 \\
 & MD3 & 2.89 & 2.24 & 4.90 & $-$8.94 \\
 & MD4 & 2.80 & 2.49 & 5.12 & $-$9.10 \\
\hline
V$_2$CO$_2$ & MD1 & 3.13 & 2.14 & 5.31 & $-$7.35 \\
 & MD2 & 2.90 & 2.38 & 4.40 & $-$9.40 \\
 & MD3 & 2.86 & 2.41 & 4.54 & $-$9.13 \\
 & MD4 & 2.82 & 2.48 & 4.70 & $-$8.78 \\
 & GS & 2.89 & 2.38 & 4.40 & $-$9.40 \\
\hline
V$_2$C(OH)$_2$ & MD1 & 3.22 & 1.78 & 7.24 & $-$7.98 \\
 & MD2 & 3.00 & 2.06 & 6.54 & $-$9.41 \\
 & MD3 & 2.93 & 2.24 & 6.77 & $-$9.02 \\
 & MD4 & 2.85 & 2.45 & 6.94 & $-$9.16 \\
\hline
V$_2$CF(OH) & MD1 & 3.21 & 1.79 & 6.26. & $-$8.39 \\
 & MD2 & 2.99 & 2.05 & 5.58 & $-$9.46 \\
 & MD3 & 2.91 & 2.24 & 5.86 & $-$8.96 \\
 & MD4 & 2.83 & 2.47 & 6.04 & $-$9.14 \\
\hline
\hline
\end{tabular}
\end{table*}
Besides, it is found that the energetically favored structures for V$_2$C-based systems functionalized with F, O and OH correspond to MD2, {where the terminal groups occupy FCC sites. However, the MD2 is dynamically unstable for V$_2$CO$_2$ as the corresponding phonon band structure presents imaginary vibrational frequencies. Therefore, among the four models, MD3 is assigned to the stable configuration for V$_2$CO$_2$. Both MD1 and MD4 are not good candidates for V$_2$CT$_2$ systems. While the ground-state structure for V$_2$CF$_2$ and V$_2$C(OH)$_2$ systems correspond to the MD2 configurations, a low-symmetry configuration (space group $P1$) is found for V$_2$CO$_2$ as the ground state (GS) structure which is close to the symmetric MD2 although with a slight modification of the atomic coordinates of both V and C atoms. This different behaviour between O-terminated system and the F/OH-terminated ones might come from the additional unpaired electron in oxygen with respect to fluorine and hydroxide groups. As symmetries in crystal structures allow an easier interpretation of both electronic and vibrational properties, the following sections only deal with pristine V$_2$C and functionalized V$_2$CF$_2$ and V$_2$C(OH)$_2$ systems.

\subsection{Electronic properties}

According to its electronic band structure, the V$_2$C mono-sheet exhibits a metallic behaviour. In contrast with the well-studied Ti$_3$C$_2$ MXene system which undercomes a metallic to indirect band gap semiconductor transition,~\cite{ad9} V$_2$CT$_2$ structures preserve their metallic character. Figure~\ref{fig:bandstructures}(a-c) shows the calculated band structures of V$_2$C, V$_2$CF$_2$ and V$_2$C(OH)$_2$, respectively, which are in agreement with previous calculations.~\cite{FNRS24,Add1,Add2} A band gap opening is only observed by J. Hu \textit{et al.},~\cite{FNRS20bis} for the terminated V$_2$CT$_2$ in the antiferromagnetic configuration. At first sight, these band structures show some similarities and, as both -F and -OH groups are able to catch only one electron from the pristine surfaces, they seem to affect the electronic band structure of V$_2$C in a similar way. However, a precise analysis of the number and origin of the bands crossing the Fermi level could provide more information regarding the electronic properties and highlight the role played by the terminal groups.
\begin{figure*}
\centering
    \includegraphics[width=\textwidth]{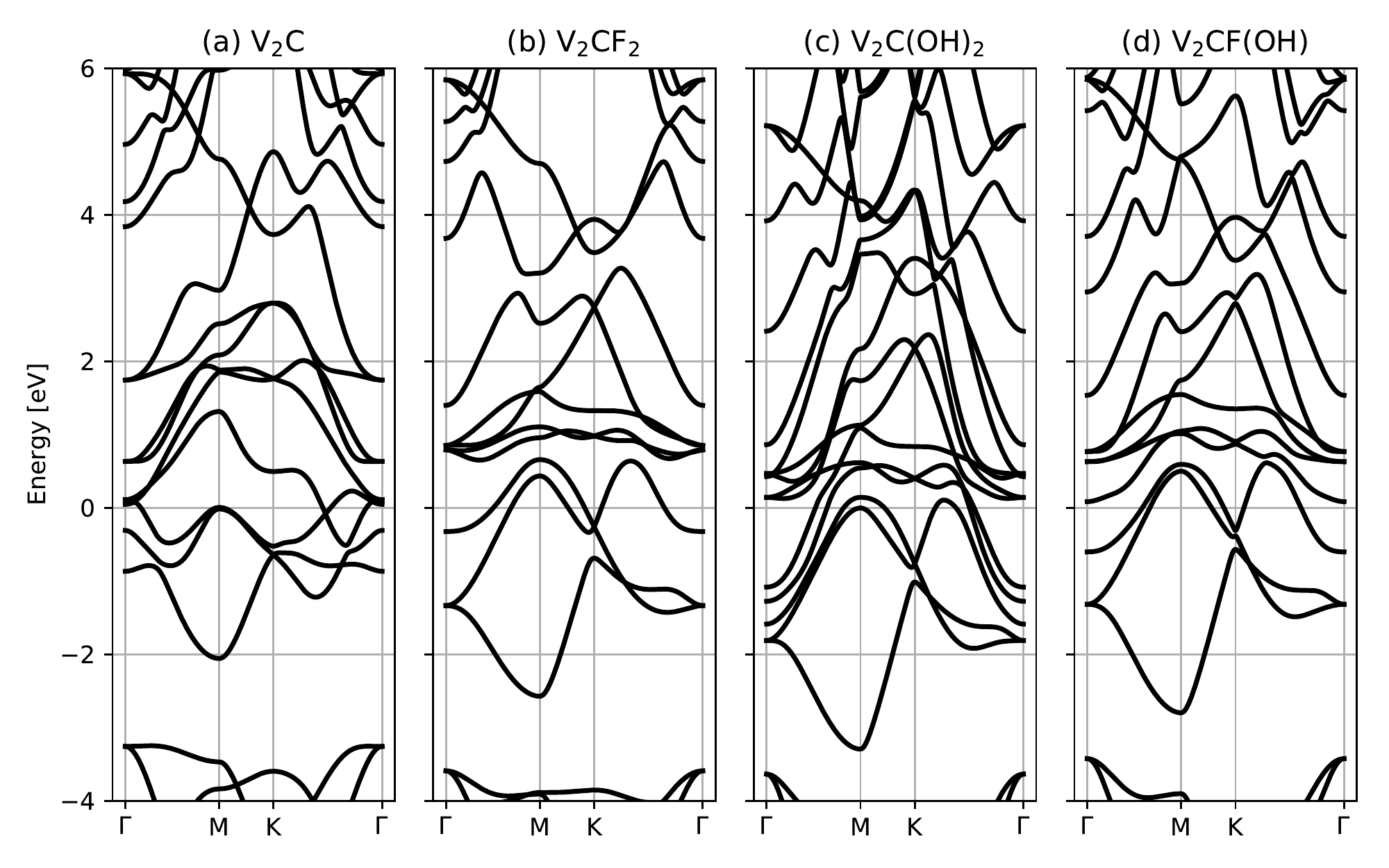}
\caption{Electronic band structures of (a) V$_2$C, (b) V$_2$CF$_2$, (c) V$_2$C(OH)$_2$, and (d) V$_2$CF(OH) in their high-symmetric configuration. The Fermi level is fixed as the reference of zero energy. \label{fig:bandstructures}}
\end{figure*}

Therefore, for both V$_2$C and V$_2$CT$_2$ (T=F,OH) systems, the local density of states (LDOS) for each atom is computed and shown in Fig.~\ref{fig:PDOS}. Moreover, in order to investigate the contribution of each atomic orbital in the LDOS, the projected densities of states (PDOSs) are also calculated.\\

When bringing atoms in close proximity, strong chemical bonds such as ionic and covalent bonds are formed, respectively, from the transfer and sharing of electrons between atoms. The two electrons involved in bonding are coupled together with a bond strength depending on the overlap between their atomic energy levels. Metal atoms usually take part in bonding via their partially-filled valence orbitals. Thereby, in V$_2$C-based systems, the C atom is involved in the bonding states through its outmost $2s$ and $2p$ orbitals, and V atoms are mainly involved through their $3d$ orbitals. In Fig.~\ref{fig:PDOS}(a), both LDOS and PDOS of pristine V$_2$C are presented and information on the hybridization of V-$3d$, C-$2p$ and C-$2s$ orbitals can be used to discuss the nature of the bonding states. In the lowest energy region, the states around 11\,eV and 13\,eV below the Fermi level can be regarded as bonding states between the C-$2s$ and the V-$3d$ orbitals. The PDOSs of C-$2p$ orbitals are mainly distributed between $-$6\,eV and $-$3\,eV. These states can clearly be assigned to bonding states between the C-$2p$ and the V-$3d$ orbitals. Around $-$3\,eV, there is a pseudogap of about 1\,eV which can be seen as the separation between the bonding and antibonding states. Finally, the states in the vicinity of the Fermi level (from $-$2\,eV and above) correspond to V-$3d$ states and are expected to give rise to electrical conductivity in the V$_2$C system [Fig.~\ref{fig:PDOS}(a)]. Upon functionalization, the atomic orbitals from the terminal groups add up in the LDOS and PDOS features. In the PDOS of V$_2$CF$_2$, the F-$2p$ orbitals are mainly located between $-$8\,eV and $-$4\,eV and can be attributed to bonding states between V-$3d$ and F-$2p$ orbitals [Fig.~\ref{fig:PDOS}(b)]. Some F-$2p$ states are also found near the Fermi energy and ensure the metallic character of V$_2$CF$_2$. In the PDOS of V$_2$C(OH)$_2$, the O-$2p$ orbitals are mainly distributed between $-$8\,eV and $-$4\,eV and can be assigned to bonding states between V-$3d$ and O-$2p$ orbitals [Fig.~\ref{fig:PDOS}(c)]. Moreover, due to the presence of hydrogen atoms, some bonding states between O-$2p$ and H-$1s$ orbitals are found in the energy range between $-$11\,eV and $-$10\,eV. In comparison with V$_2$CF$_2$, some O-$2p$ orbitals are also found around the Fermi level. The general bonding character of both V$_2$C and V$_2$CT$_2$ systems may be described as covalent-ionic and metallic. The former can be explained by the hybridization of the V-$3d$ states with the C-$2p$ and F-$2p$/O-$2p$ states presenting an important electronegativity difference, while the latter is due to the resonance of the $3d$ orbitals of the vanadium atoms around the Fermi level.~\cite{DMusic} Finally, upon adsorption of F or OH groups onto the V$_2$C surfaces, each F or OH group obtains one electron from V$_2$C leading to a slight shift of the Fermi level and of the V-$3d$, C-$2s$ and C-$2p$ bands [Figure~\ref{fig:PDOS}(b-c)]. This shift is not significant and indicates that the V-C bonds in the different V$_2$C-based systems are relatively similar, explaining also the very similar V-V distances $\sim$2.0-2.2\,\AA, (Table~\ref{tab:functionV2C}).\\
\begin{figure*}
\centering
    \includegraphics[width=\textwidth]{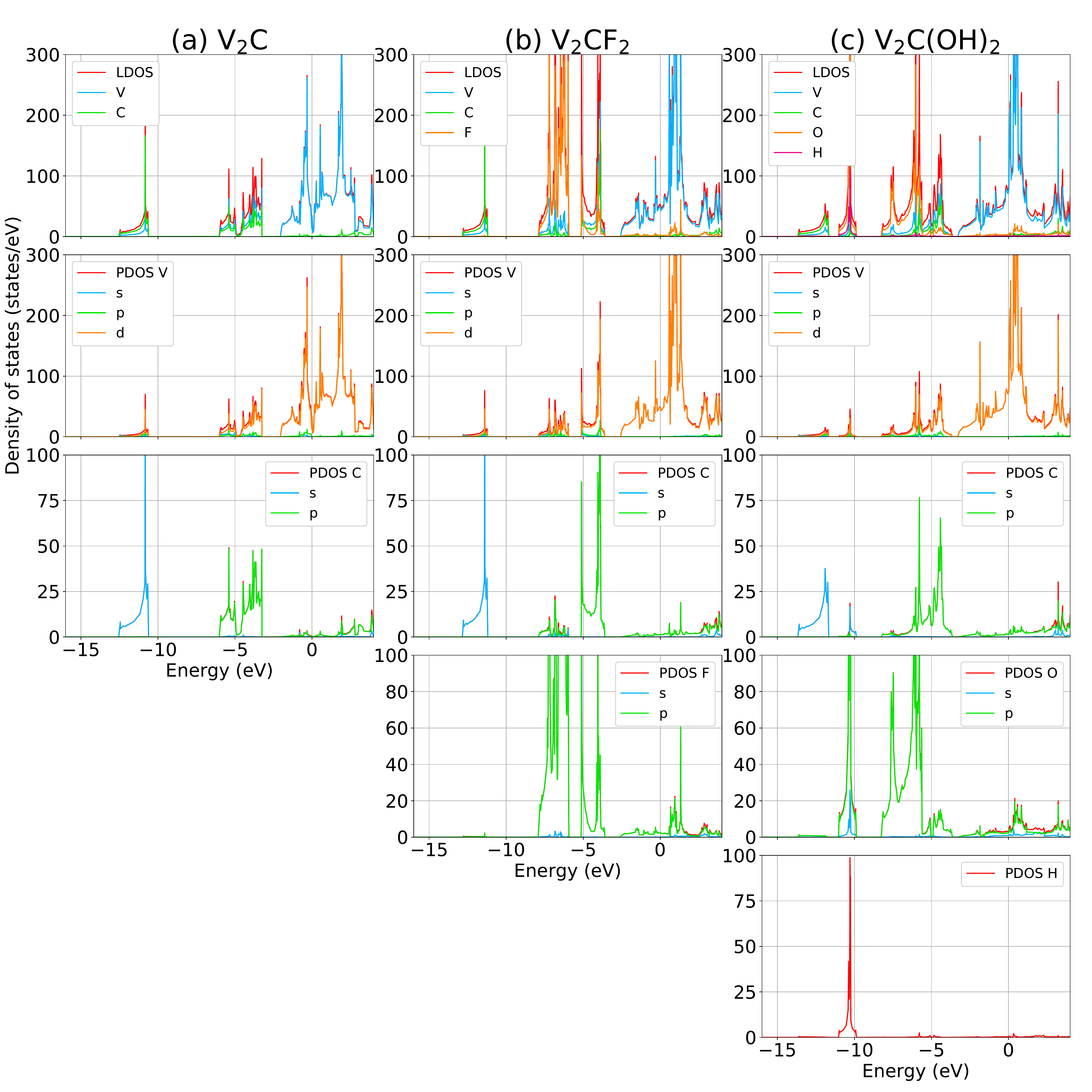}
\caption{(Color online) Local and projected density of states of (a) V$_2$C, (b) V$_2$CF$_2$ and (c) V$_2$C(OH)$_2$ in their high-symmetric configuration. The Fermi level is positioned at zero energy.\label{fig:PDOS}}
\end{figure*}

Our theoretical prediction of the metallic character of V$_2$C-based systems has been confirmed by recent electrical measurements on exfoliated V$_2$C MXene flakes.~\cite{unpub} In addition, several electronic properties, including mobilities and effective masses of the charge carriers are presently under investigation in order to reach a better understanding of the electronic transport in such systems. These results are beyond the scope of the present work and will be published in another paper.

\subsection{Vibrational analysis and Raman measurement}

According to the crystal information of the V$_2$C and V$_2$CT$_2$ systems, the 3 atoms of the primitive cell in V$_2$C mono-sheet give rise to 6 optical modes and 3 acoustic modes at the $\Gamma$ point of the Brillouin zone. In V$_2$CF$_2$, the 5 atoms of each primitive cell lead to 12 optical modes and 3 acoustic modes at the $\Gamma$ point of the Brillouin zone. In the V$_2$C(OH)$_2$, the 7 atoms of each primitive cell lead to 18 optical modes and 3 acoustic modes at the $\Gamma$ point of the Brillouin zone. The optical phonons in the center of the Brillouin zone can be classified with the following irreducible representations:
\begin{equation}
\Gamma_{\text{optical}} \left(\text{V}_2\text{C}\right) = \underbrace{E_g + A_{1g}}_{\text{Raman}} + \underbrace{A_{2u} + E_u}_{\text{IR}}
\end{equation}
\begin{equation}
\Gamma_{\text{optical}} \left(\text{V}_2\text{CF}_2\right) = \underbrace{2E_g + 2A_{1g}}_{\text{Raman}} + \underbrace{2A_{2u} + 2E_u}_{\text{IR}}
\end{equation}
\begin{equation}
\Gamma_{\text{optical}} \left(\text{V}_2\text{C(OH)}_2\right) = \underbrace{3E_g + 3A_{1g}}_{\text{Raman}} + \underbrace{3A_{2u} + 3E_u}_{\text{IR}}
\end{equation}
The vibrational frequencies of the symmetric structures of V$_2$C, V$_2$CF$_2$ and V$_2$C(OH)$_2$ are listed in Table~\ref{modesV2C}.
\begin{table*}
\caption{Vibrational modes frequencies of the stable V$_2$C-based mono-sheets. $E_{g}$ and $E_{u}$ modes are doubly degenerate.\label{modesV2C}}
\begin{tabular}{l@{\hskip 0.5cm} c@{\hskip 0.3cm} c@{\hskip 0.3cm} c@{\hskip 0.3cm} c@{\hskip 0.3cm} c@{\hskip 0.3cm} c}
\hline
\hline
Raman modes & $E_g$ & $A_{1g}$ & $E_{g}$ & $A_{1g}$ & $E_g$ & $A_{1g}$ \\
\hline
V$_2$C & 224.4 & 358.9 & - & - & - & - \\
V$_2$CF$_2$ & 197.8 & 289.6 & 281.6 & 525.6 & - & - \\
V$_2$C(OH)$_2$ & 214.3 & 301.1 & 296.6 & 530.2 & 443.2 & 3613.3 \\
\hline
IR modes & $A_{2u}$ & $E_u$ & $E_u$ & $A_{2u}$ & $E_u$ & $A_{2u}$ \\
\hline
V$_2$C & 610.4 & 700.8 & - & - & - & - \\
V$_2$CF$_2$ & 636.0 & 761.0 & 221.3 & 442.8 & - & - \\
V$_2$C(OH)$_2$ & 627.4 & 757.5 & 285.5 & 446.7 & 435.7 & 3600.0 \\
\hline
\hline
\end{tabular}
\end{table*}
The correspondence between the vibrational modes of the three systems is determined according to the direction of the vibrations and the nature of the contributing atoms. Schematic representations of the Raman active modes of the pristine V$_2$C are illustrated in Fig.~\ref{Vibration}.\\
\begin{figure*}
 \includegraphics[width=.5\textwidth]{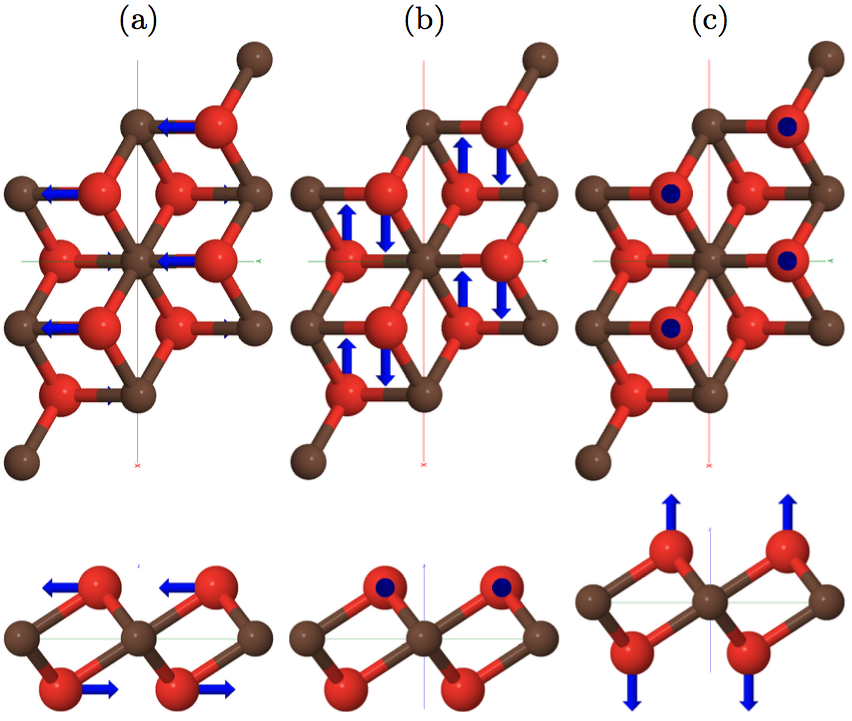}
\caption{(Color online) Schematic representations of the Raman active modes of pristine V$_2$C MXene: (a-b) in-plane vibration of V atoms at 224.4\,cm$^{-1}$ corresponding to $E_g$ mode and (c) out-of-plane vibration of V atoms at 358.9\,cm$^{-1}$ corresponding to $A_{1g}$ mode. Vanadium (V) atoms are in red (gray) and Carbon (C) atoms are in brown (dark gray). \label{Vibration}}
\end{figure*}

The knowledge of the phonon spectrum of a material is essential for the understanding of a wide set of macroscopic properties, including the electrical and thermal conductivities. The phonon spectrum of V$_2$C was recently reported by Gao \textit{et al.},~\cite{Add1} highlighting the stability of the system. Phonon dispersions and phonon densities of states (PhDOS) for the V$_2$C, V$_2$CF$_2$ and V$_2$C(OH)$_2$ mono-sheets are presented in Fig.~\ref{fig:PhononBS}(a-c).
\begin{figure*}
\centering
\begin{tabular}{c c}
(a) V$_2$C & (b) V$_2$CF$_2$ \\
    \includegraphics[width=.48\textwidth]{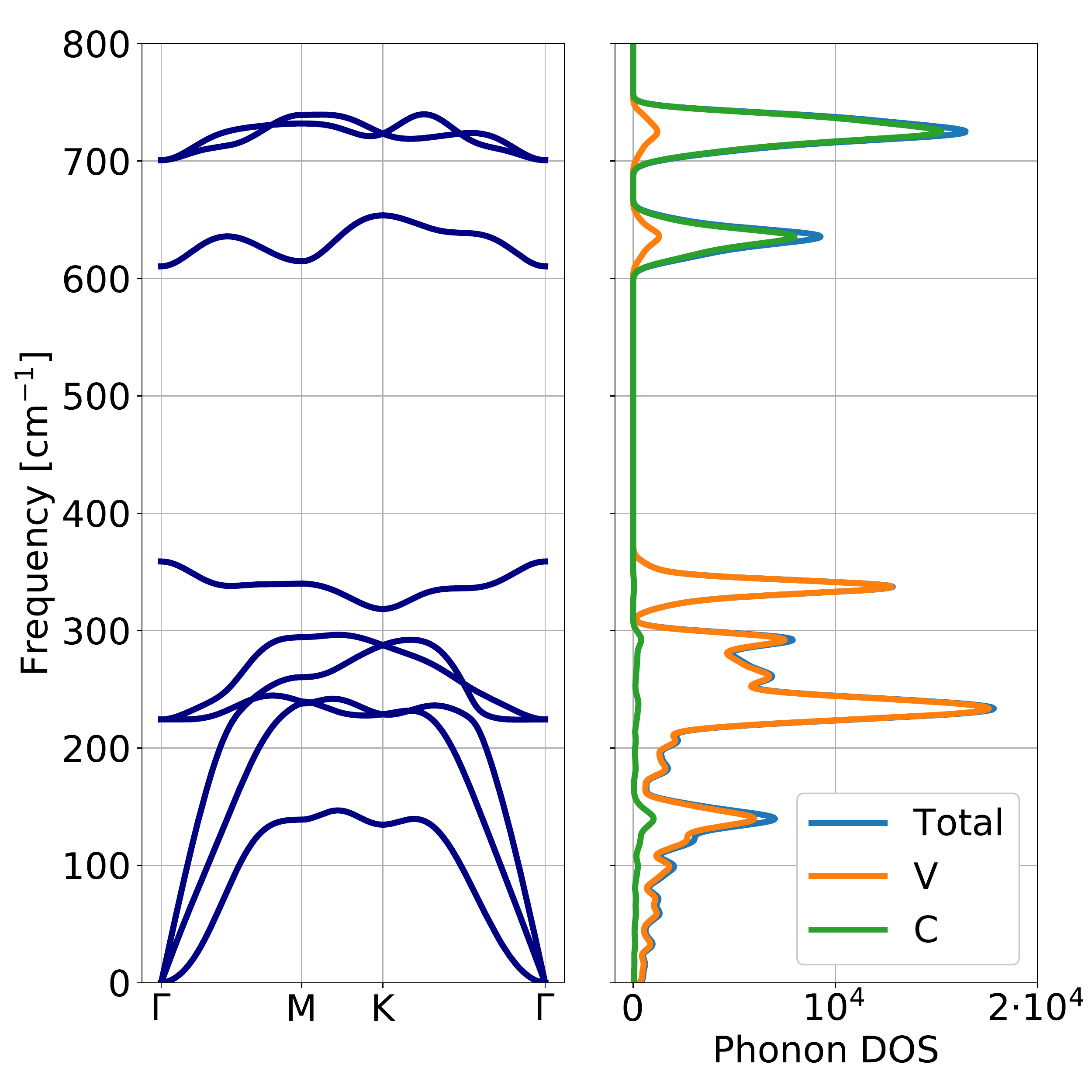} &
    \includegraphics[width=.48\textwidth]{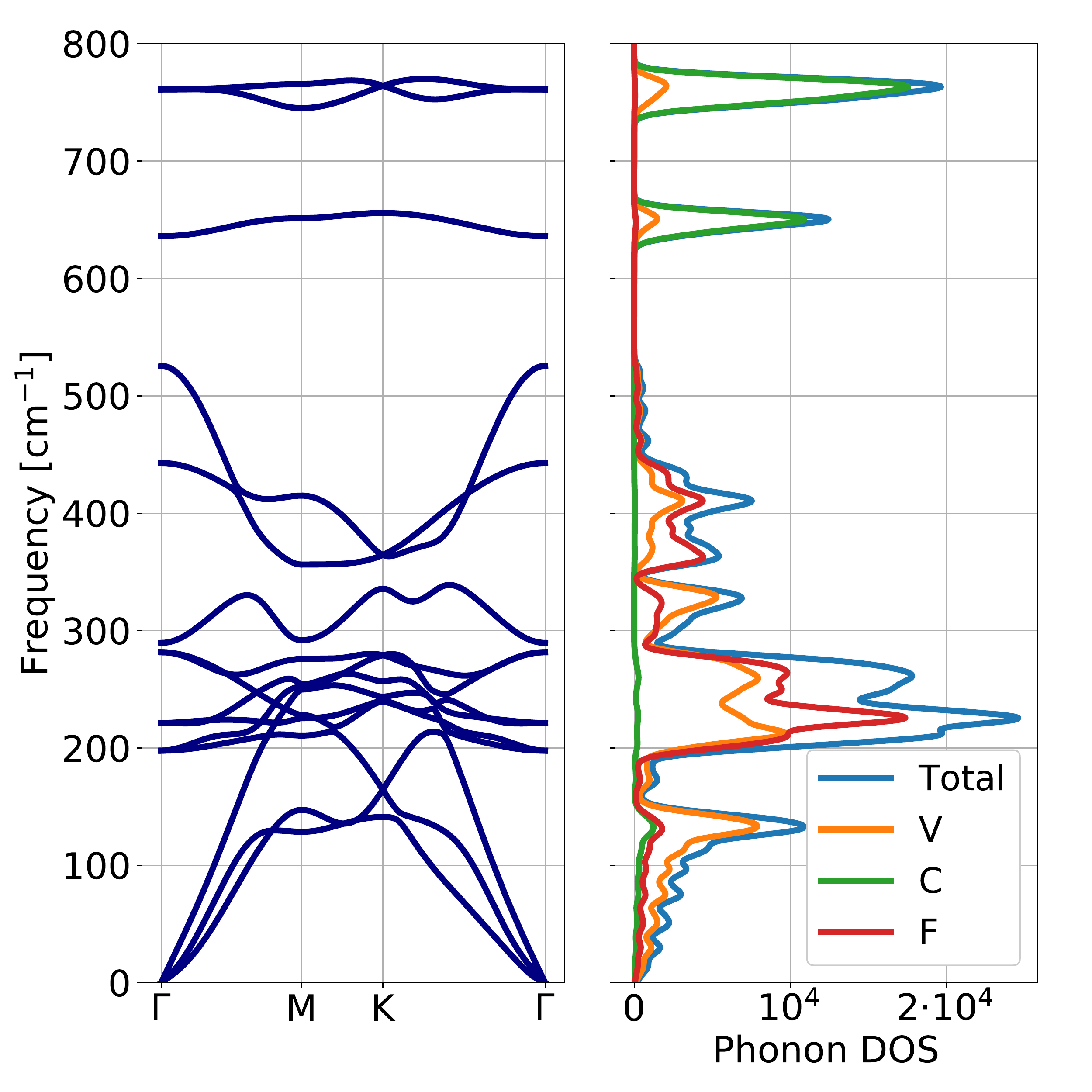} \\
(c) V$_2$C(OH)$_2$ & (d) V$_2$CF(OH) \\
    \includegraphics[width=.48\textwidth]{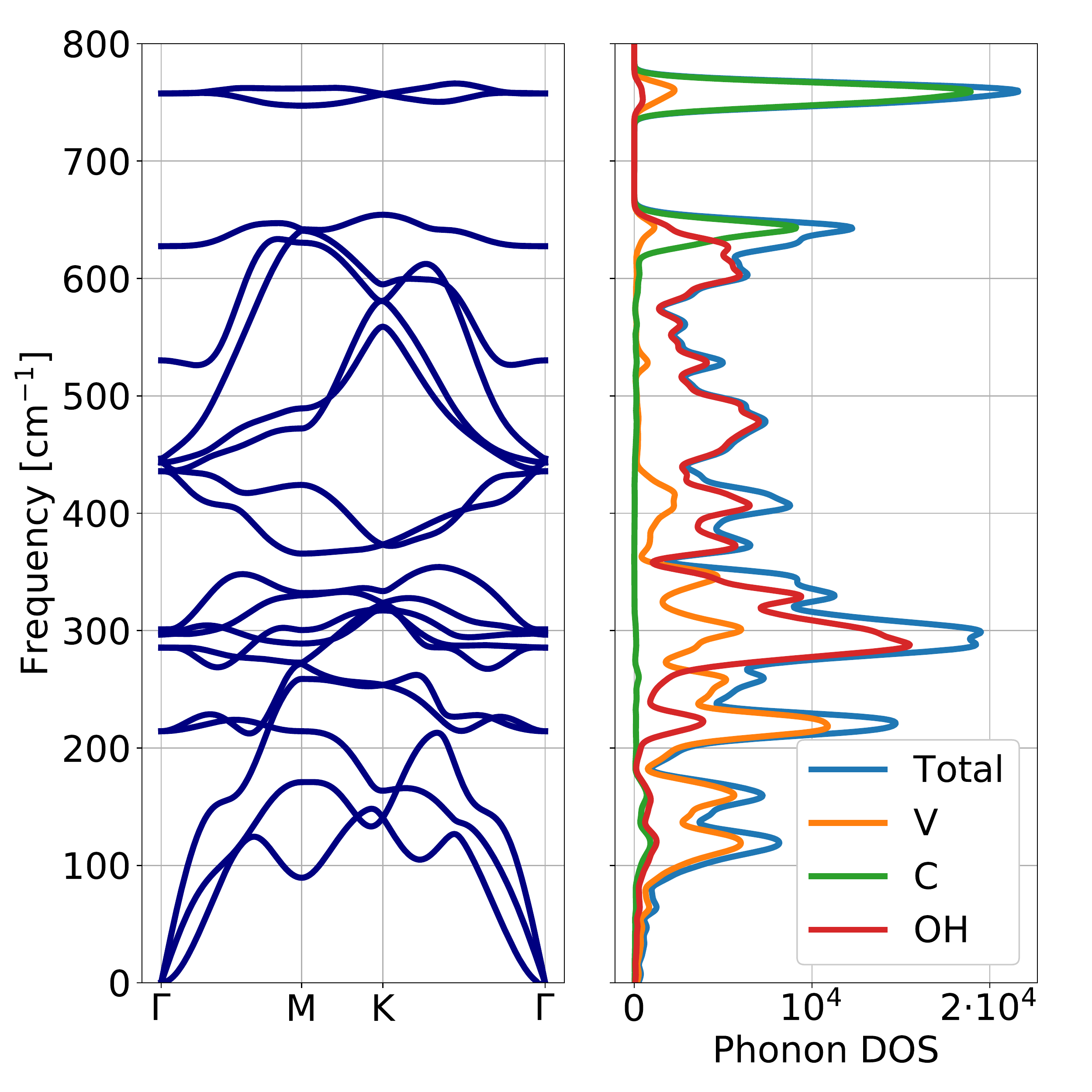} &
    \includegraphics[width=.48\textwidth]{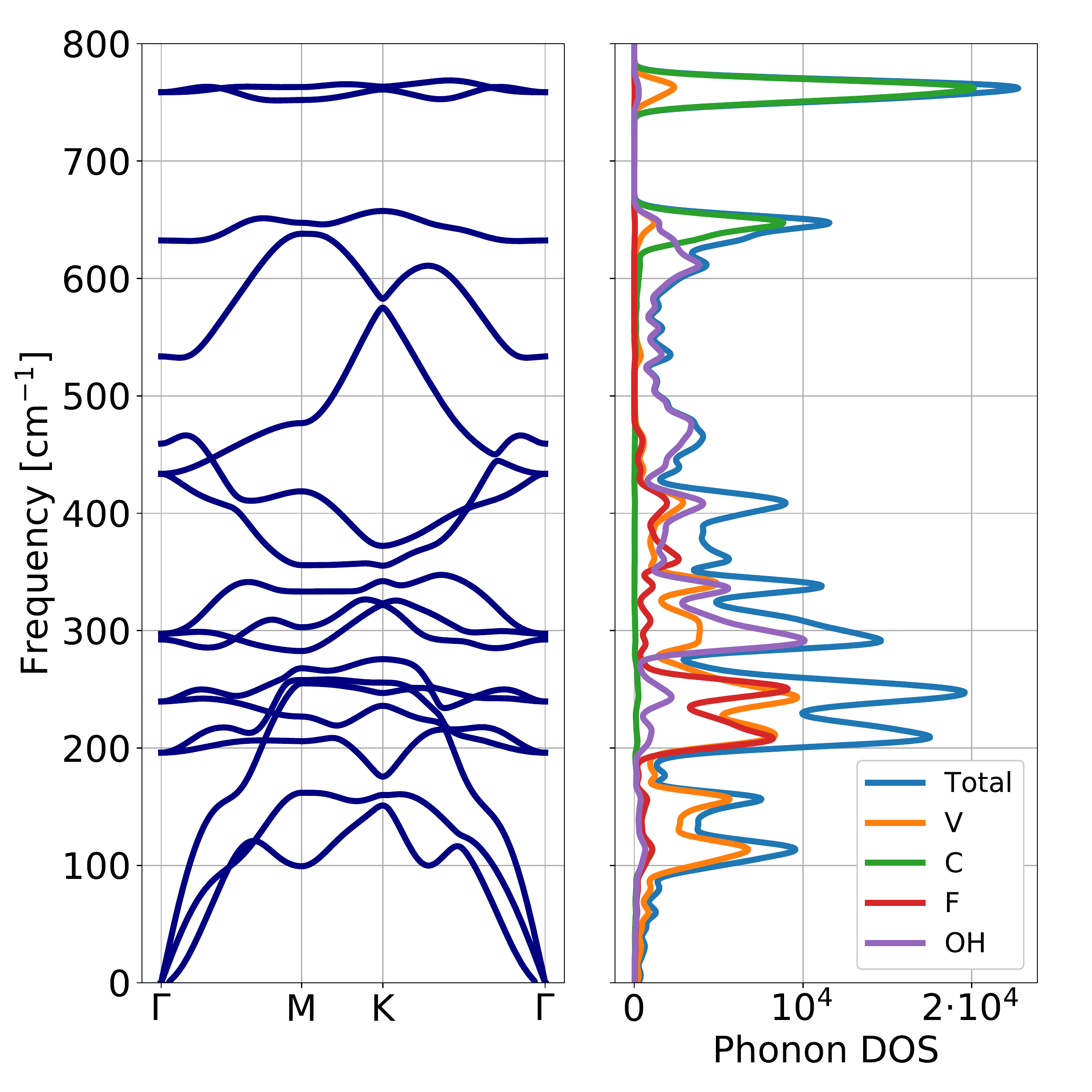} \\
\end{tabular}
\caption{(Color online) Phonon band structures and densities of states of (a) V$_2$C, (b) V$_2$CF$_2$, (c) V$_2$C(OH)$_2$ and (d) V$_2$CF(OH) in their high-symmetric nonmagnetic configuration. For the three systems, the phonon density of states for each atom is also plotted allowing for a better visualization of the contributing atoms in each normal mode. \label{fig:PhononBS}}
\end{figure*}
These phonon band structures are plotted along the path $\Gamma (0, 0, 0) \to M (1/2,0, 0) \to K (1/3, 1/3, 0) \to \Gamma (0, 0, 0)$. The phonon dispersions have three acoustic modes. Two of them exhibit a linear dispersion near $\Gamma$ and correspond to in-plane rigid-body motions. In contrast, the third acoustic mode corresponding to out-of-plane vibration has a quadratic dispersion close to $\Gamma$ and a lower energy in the rest of the spectrum. This quadratic dependence is analogous to the one observed in graphene,~\cite{ad15} and in MoS$_2$.~\cite{ad16} In Fig.~\ref{fig:PhononBS}(a), the vibrations of the V$_2$C system are below 750\,cm$^{-1}$. The highest peaks in energy in the PhDOS corresponding to the vibration of carbon atoms have a large intensity due to the flatness of the bands around $\Gamma$. Moreover, the first three optical branches in the phonon spectrum demonstrate significantly lower frequencies close to the three acoustic phonon branches, thus inducing a phonon gap of about 250\,cm$^{-1}$ between the lower three and the upper three optical branches. This is a general property of MXenes and has already been observed previously in several MXene systems.~\cite{FNRS16} For the functionalized forms, this phonon gap between optical phonons is filled by additional optical phonon branches. This makes the most noticeable difference between pristine and terminated MXene systems. The phonon spectra of the terminated V$_2$CT$_2$ systems [Fig.~\ref{fig:PhononBS}(b-c)] reveal that the highest optical phonon frequencies are between 750\,cm$^{-1}$ and 800\,cm$^{-1}$, slightly larger than for the pristine case.\\

To further understand the role played by the surface functionalization in the MXene layer and its influence on the vibrational properties, the Raman and infrared (IR) active frequencies of the three mono-sheets are compared.

First of all, as presented in Table~\ref{modesV2C}, all terminal atoms strongly influence the normal modes frequencies calculated for pristine V$_2$C. Regarding the Raman active frequencies, the doubly-degenerated $E_g$ modes at 224\,cm$^{-1}$ and the $A_{1g}$ mode at 359\,cm$^{-1}$ in the pristine V$_2$C mono-sheet correspond to in-plane vibration [Fig.~\ref{Vibration}(a-b)] and out-of plane vibration of V atoms [Fig.~\ref{Vibration}(c)], respectively. The $E_g$ frequency shifts to lower wavenumbers upon terminating with F and OH functional groups. The $A_{1g}$ mode is softened to 290 and 301\,cm$^{-1}$ upon functionalization with -F and -OH groups. Since their main contributions are from vibrations of V atoms, it can be inferred that the vibrations of the terminal atoms weaken the motion of the V atoms. The IR active frequencies located at 610 and 701\,cm$^{-1}$ correspond to out-of-plane motions of C atoms ($A_{2u}$ mode) and in-plane motion of C atoms (doubly degenerated $E_u$ modes), respectively. The former mode at 610\,cm$^{-1}$ shifts to 636 and 627\,cm$^{-1}$ while the latter one at 701\,cm$^{-1}$ is shifted to 761 and 758\,cm$^{-1}$ when terminating with F and OH groups, respectively. Therefore, the functional groups tend to strengthen both in-plane and out-of-plane motions of C atoms.

The remaining modes are attributed to the in-plane and out-of-plane vibrations of -F and -OH groups. The phonon mode at about 3600\,cm$^{-1}$ in the V$_2$C(OH)$_2$ mono-sheet is dominated by the vibration of the -OH terminal groups. In V$_2$CF$_2$ mono-sheet, the frequencies around 500\,cm$^{-1}$ are due to collaborative vibrations of V and F atoms. Collaborative vibrations of V and OH are also observed around 500\,cm$^{-1}$ in V$_2$C(OH)$_2$. As the terminal -F and -OH groups strongly bond with V, the band gap centered around 500\,cm$^{-1}$ in the V$_2$C mono-sheet disappears in V$_2$CT$_2$ (T = F, OH). Lower-frequency phonons are due to collaborative motion of all atoms in the mono-sheets [Fig.~\ref{fig:PhononBS}(a-c)].\\

Thanks to its sensitivity to very small changes in crystal structures, Raman spectroscopy is used for characterizing the composition and the quality of samples. However, before it can be used for this purpose, it is fundamental to properly define the peaks positions and to assign them with the phonons computed at the Brillouin zone center ($\Gamma$ point). This allows to identify the contribution of each atom and group of atoms in the various vibrational modes corresponding to the reported peaks. The Raman spectrum of V$_2$C-based samples simply deposited on a glass slide is presented in Fig.~\ref{fig:SEM}(b). The comparison between the Raman spectrum of the 3D V$_2$AlC MAX phase and the corresponding 2D V$_2$C MXene system indicates a global reduction of the peak intensities and a broadening of the peaks, probably due to the larger interlayer spacing in the 2D system, with respect to its 3D counterpart. Globally speaking, the two Raman spectra are very different, with sharp and well-defined peaks in the Raman spectrum of V$_2$AlC in contrast with those observed in the Raman spectrum of V$_2$C. Four peaks are identified in the V$_2$AlC spectrum at 158\,cm$^{-1}$ ($E_{2g}$), 239\,cm$^{-1}$ ($E_{2g}$), 258\,cm$^{-1}$ ($E_{1g}$), and 360\,cm$^{-1}$ ($A_{1g}$).~\cite{V2AlCMAX} The first two peaks, corresponding to in-plane vibrations of V and Al atoms, seem to completely disappear, due to the removal of Al atoms, as confirmed by our EDS measurements.~\cite{luunp} The other two peaks, corresponding respectively to in-plane and out-of-plane vibrations of V atoms, can still find an equivalent in the V$_2$C spectrum, as confirmed by our theoretical prediction of an $E_g$ mode around 224\,cm$^{-1}$ and an $A_{1g}$ mode around 359\,cm$^{-1}$ for the V$_2$C system. Besides, as the etching process involves the substitution of Al atoms with lighter elements (F, OH), additional peaks can be observed at higher frequencies. These features make us confident with the achievement of the etching process and the resulting removal of the Al atoms from the V$_2$AlC sample. The calculated Raman active frequencies of both V$_2$C and V$_2$CT$_2$ systems are reported in Table~\ref{modesV2C} and also illustrated in Fig.~\ref{fig:Raman}. The predicted peaks positions for the V$_2$C mono-layer are not sufficient to describe the experimental Raman spectrum, especially above 400\,cm$^{-1}$. Upon termination with -F and -OH groups, additional Raman frequencies appear around 525\,cm$^{-1}$ and around 443, 530 and 3613\,cm$^{-1}$, respectively. These frequencies roughly match most of the bands in the collected spectrum and it comes up that the peaks around 430 and 530\,cm$^{-1}$ are clearly indicative of terminated V$_2$CT$_2$ MXene. The peak around 3600\,cm$^{-1}$ in the experimental spectrum confirms the presence of -OH terminal groups at the sample surface. However, there are still some discrepancies between the theoretical predictions and the experimental Raman frequencies, as some predicted frequencies do not correspond to any Raman bands, and inversely. More specifically, the theoretical simulations on both pristine and terminated V$_2$C MXenes are not able to describe the hump between 600 and 700\,cm$^{-1}$. Different factors can explain these deviations. First, the calculations are performed for homogeneous terminal groups, while the surface of our exfoliated sample is randomly terminated with both -F and -OH groups.~\cite{FNRS22} Second, the calculations deal with flat mono-layers, while our sample consists in a stacking of sheets including stress and corrugations [Fig.~\ref{fig:SEM}(a)]. For instance, the presence of intrinsic defects such as vacancies and adatoms was evidence in Ti$_3$C$_2$T$_x$ systems.~\cite{ad6}\\
\begin{figure*}
\centering
\begin{tabular}{c c}
\includegraphics[width=0.4\textwidth]{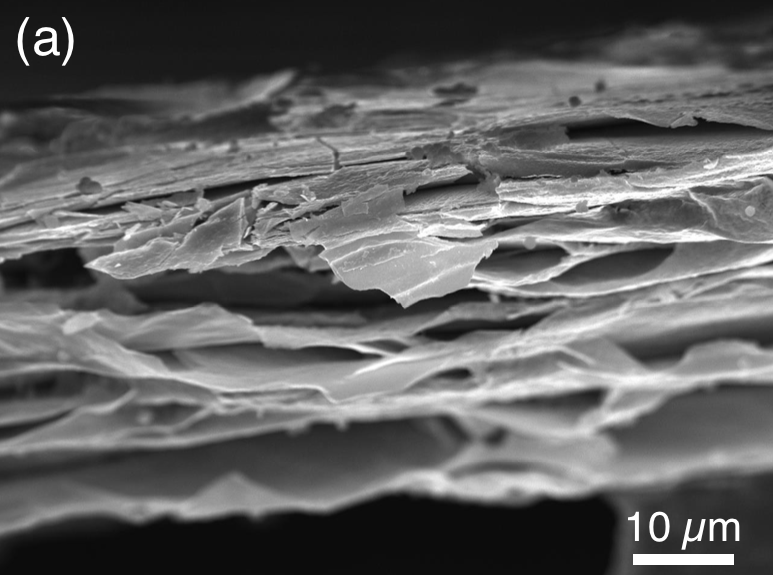} &
\includegraphics[width=0.54\textwidth]{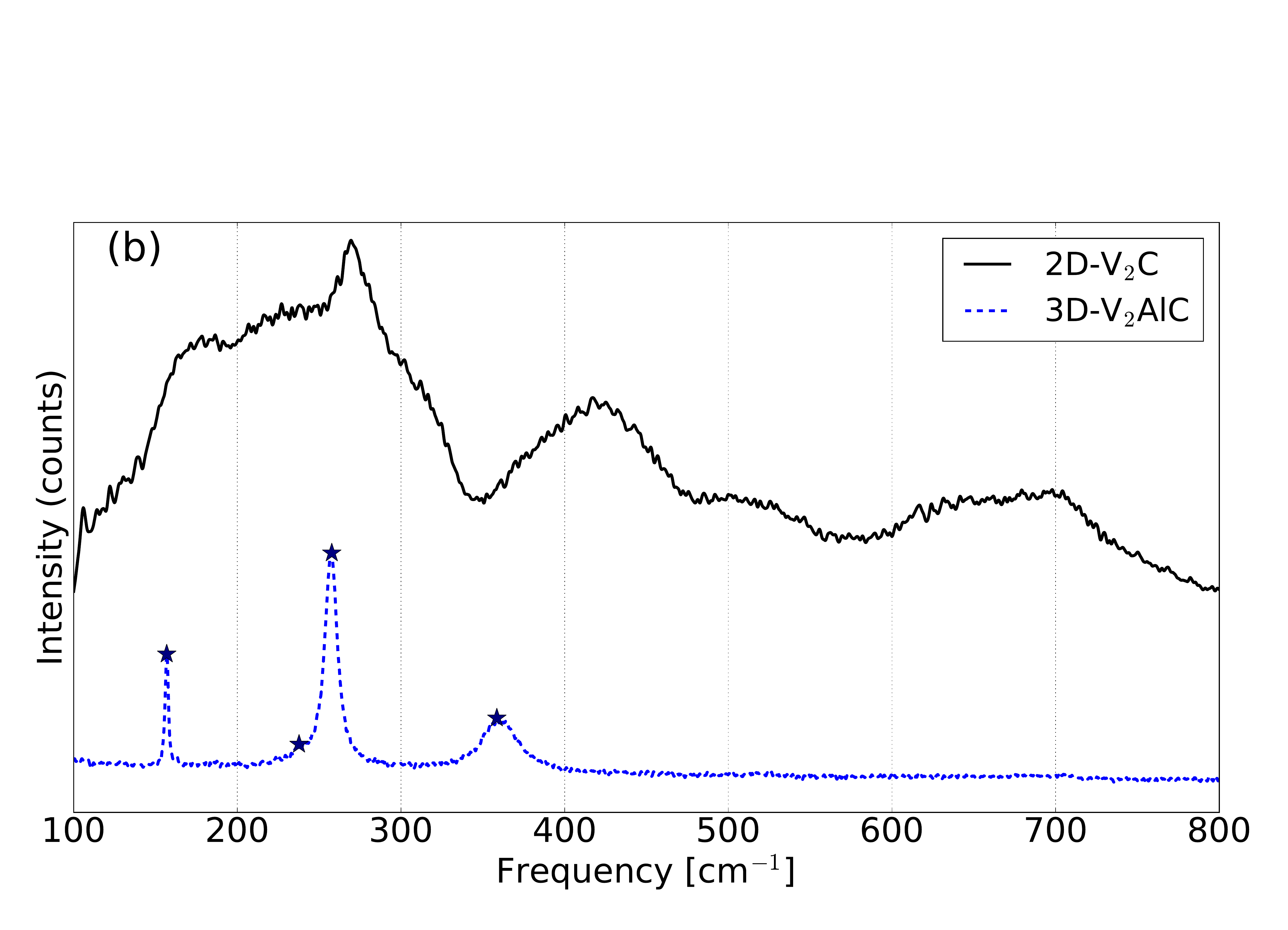}\\
\end{tabular}
\caption{(Color online) (a) Scanning Electron Microscopy (SEM) image and (b) Raman spectrum of the 3D MAX V$_2$AlC and the 2D MXene produced by HF-etching of V$_2$AlC sample. Note the sample consists in the stacking of numerous 2D layers, with curved paper-like morphology.\label{fig:SEM}}
\end{figure*}
\begin{figure*}
\centering
\includegraphics[width=\textwidth]{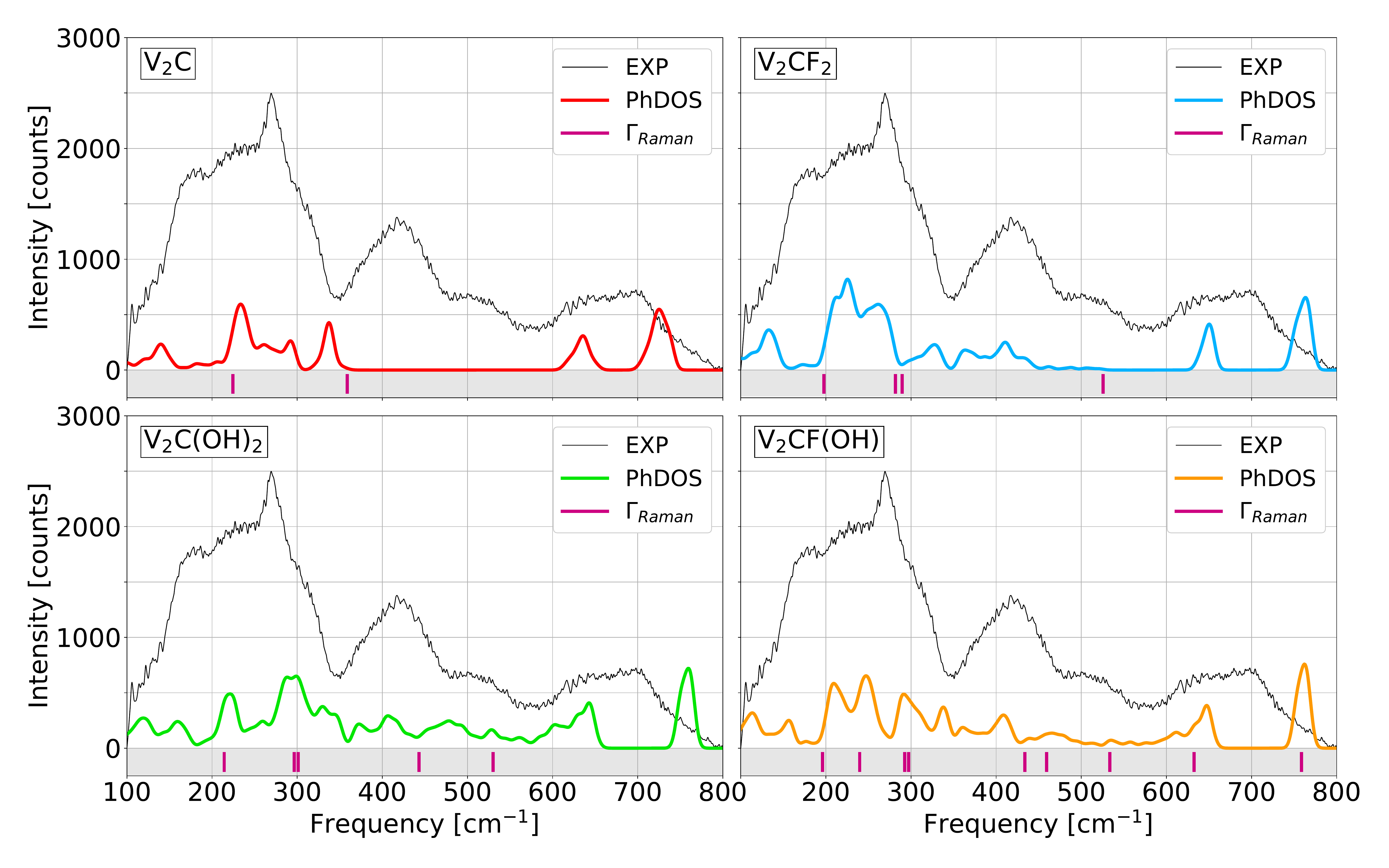}
\caption{(Color online) Raman spectrum of the exfoliated V$_2$C-based sample collected at room temperature. The calculated Raman-active frequencies and the total phonon densities of states of the V$_2$C, V$_2$CF$_2$, V$_2$C(OH)$_2$ and V$_2$CF(OH) mono-sheets are also included under the experimental spectrum for comparison. The matching between the predicted normal mode frequencies and the experimental spectrum confirms the presence of heterogeneous terminal groups randomly distributed at the V$_2$C surface. \label{fig:Raman}}
\end{figure*}

With the hope to improve the agreement between the experiment and the theoretical predictions, we investigate the effect of heterogeneous termination of the Raman active modes. Using the same principles as for the homogeneous functionalization, the V$_2$CF(OH) mono-sheet (F terminated on one side while OH terminated on the other side of the mono-sheet) is built. The four models (MD1-4) are tested and it is found that an equivalent of MD2 is the most energetically favorable configuration (Table~\ref{tab:functionV2C}). As presented in Fig.~\ref{fig:bandstructures}(d), V$_2$CF(OH) is still electrically conductive. The heterogeneity of the terminal groups induces a difference in the V-T bonding lengths on both sides of the MXene layer. As the reduced coordinates are modified, the Wyckoff positions are influenced and, hence, the space group is moved to $P3m1$ (156). For such configuration, the irreducible representation is given by 
\begin{equation}
\Gamma_{\text{optical}} \left(\text{V}_2\text{CF(OH)}\right) = \underbrace{5A_1 + 5E}_{\text{Raman+IR}}
\end{equation}
for V$_2$CF(OH). The normal mode frequencies for this system are listed in Table~\ref{tab:RamanMix} and compared to the experimentally collected Raman spectrum in Fig.~\ref{fig:Raman}(d).
\begin{table*}
\small
\caption{Raman and infrared (IR) active vibrational modes of stable V$_2$C-based systems with heterogeneous functional groups. All the vibrational modes are Raman and IR active.\label{tab:RamanMix}}
\begin{tabular}{l@{\hskip 0.5cm} c@{\hskip 0.5cm} c@{\hskip 0.3cm} c@{\hskip 0.3cm} c@{\hskip 0.3cm} c@{\hskip 0.3cm} c@{\hskip 0.3cm} c@{\hskip 0.3cm} c@{\hskip 0.3cm} c@{\hskip 0.3cm} c@{\hskip 0.3cm} c}
\hline
\hline
 & MD & $E$ & $E$ & $A_1$ & $E$ & $E$ & $A_1$ & $A_1$ & $A_1$ & $E$ & $A_1$ \\
\hline
V$_2$CFO & MD2 & 188.1 & 214.7 & 290.7 & 394.7 & 497.5 & 584.3 & - & 704.0 & 723.0 & - \\
V$_2$CF(OH) & MD2 & 196.0 & 239.7 & 292.5 & 297.3 & 433.6 & 459.3 & 533.6 & 632.4 & 758.7 & 3616.6 \\
V$_2$CO(OH) & MD2 & 202.9 & 227.4 & 293.7 & 342.7 & 381.9 & 531.1 & 573.5 & 696.6 & 696.7 & 3652.6\\
\hline
\hline
\end{tabular}
\end{table*}

There are four main additional Raman active modes at 434\,cm$^{-1}$, 459\,cm$^{-1}$, 632\,cm$^{-1}$ and 759\,cm$^{-1}$, with respect to the homogeneous situations. The first two peaks match well with the band around 430\,cm$^{-1}$ and the last two peaks with the hump centered at 650\,cm$^{-1}$ in the experimental spectrum. This close matching between the theoretically predicted positions of the Raman active peaks and the experimental Raman spectrum reflects the fact that the terminal groups are actually heterogeneous and most probably randomly distributed at the V$_2$C surface.\\

Table~\ref{tab:summary} summarizes the assignment of the experimental peaks with the theoretical predictions. These results suggest that the presence of Raman peaks between 400\,cm$^{-1}$ and 550\,cm$^{-1}$ is indicative of terminated V$_2$CT$_2$ systems, while the peak centered at 650\,cm$^{-1}$ seems to originate from the presence of heterogeneous functional groups at the sample surface. The presence of -OH groups gives rise to a peak around 3600\,cm$^{-1}$ in the experimental spectrum. The deviations between the computed and collected Raman frequencies are reported in Table~\ref{tab:summary} and indicate that the experimental Raman frequencies are correctly predicted, with a maximal deviation of 15\% and 3\% in the low- and high- frequency range, respectively.

\begin{table*}
\small
\caption{Experimental Raman peaks positions compared to the predicted Raman active frequencies computed for V$_2$C, V$_2$CF$_2$, V$_2$C(OH)$_2$ and V$_2$CF(OH) mono-sheets. The last row corresponds to the maximal deviation (in \%) of the computed frequencies for V$_2$CF(OH) with respect to the experimental data.\label{tab:summary}}
\begin{tabular}{l@{\hskip 0.5cm} c@{\hskip 0.3cm} c@{\hskip 0.3cm} c@{\hskip 0.3cm} c@{\hskip 0.3cm} c@{\hskip 0.3cm} c@{\hskip 0.3cm} c}
\hline
\hline
 & \multicolumn{7}{c}{Raman frequencies [cm$^{-1}$]}\\
\hline
EXP spectrum & $\sim$170 & $\sim$230 & $\sim$270 & $\sim$430 & $\sim$540 & $\sim$650 & $\sim$3600\\
\hline
V$_2$C & - & 224 & 359 & - & - & - & -\\
V$_2$CF$_2$ & - & 198 & 282, 290 & - & 526 & - & -\\
V$_2$C(OH)$_2$ & - & 214 & 297, 301 & 443 & 530 & - & 3613\\
V$_2$CF(OH) & 196 & 240 & 293, 297 & 434, 459 & 534 & 632 & 3617\\
$\Delta_{\text{MAX}}$(\%) & 15 & 4 & 10 & 7 & 1 & 3 & 0.5\\
\hline
\hline
\end{tabular}
\end{table*}

\section{\label{sec:level3} Conclusions}
In conclusion, the static and dynamical properties of pristine bare V$_2$C and terminated V$_2$CT$_2$ (T = F, O and OH) mono-sheets have been investigated using first-principles techniques. This work  gives insight on the structural, electronic and vibrational properties of an emerging 2D material. The optimized crystal structure of V$_2$C is dynamically stable. It is also found that the surface functionalization with F and OH groups is energetically favorable, in opposition to oxygen termination. Regarding the electronic properties, both V$_2$C and V$_2$CT$_2$ (T = F, OH) systems are metallic. The electronic bands crossing the Fermi level come up from V-$3d$ orbitals. The analysis of the phonon dispersion and PDOS of the bare V$_2$C mono-sheet demonstrates the presence of a phonon gap between the lowest three and the three upper most optical branches. Upon functionalization, such phonon band gap disappears as the gap is bridged with vibrations of the terminal groups (F and OH). It is shown that the terminal atoms weaken the motions in which the V atoms are involved, while strengthening the vibrations of the C atoms. The corresponding vibrational frequencies are drastically changed when the nature of the functional group is modified. Vibrational properties were computed with the aim to understand and characterize the origin of the Raman active peaks in the experimental spectrum. A first attempt of assignment of the atomic motions with the experimental Raman peaks is proposed.  The comparison of our theoretical predictions with the experimental Raman spectrum gives a satisfying agreement, as both the number of peaks and their positions are well-described, especially when considering mixed terminal groups at the MXene's surface. The list of central peaks frequencies for each hump in the experimental spectrum reported in Table~\ref{tab:summary}, can thus be considered as a signature of V$_2$C samples with different chemical terminations. This work is a prerequisite to experimental nondestructive identification and synthesis of V$_2$C-based 2D materials.

\begin{acknowledgments}
This work was supported by the UCL through a FSR Grant (A.C.) and by the FRS-FNRS through a FRIA Grant (L.S.). The authors acknowledge financial support from the F\'ed\'eration Wallonie-Bruxelles through the Action de Recherche Concert\'ee (ARC) on 3D nanoarchitecturing of 2D crystals (N$\degree$16/21-077), from the European Union's Horizon 2020 researchers and innovation programme (N$\degree$696656), and from the Belgium FNRS. The authors are also indebted to the Flag-ERA JTC 2017 project entitled "MORE-MXenes". Computational resources were provided by the supercomputing facilities of the Universit\'e catholique de Louvain (CISM/UCL) and the Consortium des Equipements de Calcul Intensif en F\'ed\'eration Wallonie Bruxelles (CECI) funded by the Fonds de la Recherche Scientifique de Belgique (F.R.S.-FNRS) under convention N$\degree$2.5020.11. Crystal growth was financially supported by the Agence Nationale de la Recherche through the project ANR-13-BS09-0024. All structural models were rendered using VESTA~\cite{vesta} and Jmol.~\cite{jmol}
\end{acknowledgments}

%% Create the reference section using BibTeX:
%\bibliography{manuscript}
%

\end{document}